\documentclass[preprint,10p+t]{elsarticle}

\usepackage{amssymb}
\usepackage{float}
\usepackage{amsmath}
\usepackage{setspace}
\usepackage[labelfont=bf]{caption} 
\usepackage{graphicx}
\usepackage{xcolor}
\usepackage{svg}

\journal{Journal of Electronic Materials}
\begin{document}
\begin{frontmatter}

\title{Optimizing the Optical Properties of Tin Oxide Aerogels through Defect Passivation}

\author[label1,label2]{John F. Hardy II\corref{cor1}}
\author[label1,label3]{Madison King}
\author[label1,label3]{Stephanie Hurst}
\author[label1,label2]{Carlo R. daCunha}

\affiliation[label1]{organization={Center for Materials Interfaces in Research and Applications, Northern Arizona University},
            addressline={700 Osborn Dr.}, 
            city={Flagstaff},
            postcode={86011}, 
            state={AZ},
            country={USA}}
\cortext[cor1]{Corresponding author. Email: jfh67@nau.edu}
\affiliation[label2]{organization={School of Informatics, Computer, and Cyber Systems, Northern Arizona University},addressline={1295 Knoles Drive},city={Flagstaff},postcode={86011},state={AZ},country={USA}}

\affiliation[label3]{organization={Department of Chemistry and Biochemistry, Northern Arizona University},addressline={700 Osborn Dr.},city={Flagstaff},postcode={86011},state={AZ},country={USA}}

\begin{abstract}
    Tin oxide aerogels were synthesized using an epoxide-assisted technique and characterized with Fourier transform infrared, X-ray diffraction, and UV-Vis to study the effects of post-synthesis annealing and peroxide treatment. While bulk tin oxide exhibits an optical bandgap of $3.6$ eV, its aerogel form often displays a larger apparent bandgap around $4.6$ eV due to defects. Our study reveals that annealing induces a partial phase change from SnO$_2$ to SnO, but is ineffective in removing defects. Conversely, peroxide passivation effectively lowers the bandgap and disorder levels, suggesting that dangling bonds are the primary cause of the increased bandgap in tin oxide aerogels. These findings offer insights for optimizing the optical properties of tin oxide aerogels for applications like solar cells.
\end{abstract}


\end{frontmatter}
\doublespacing

\section{Introduction}
Tin oxide (SnO$_2$) has received attention in recent years due various applications, such as gas sensing \cite{SnO2Gas}, catalysis \cite{SnO2Cat}, and energy storage \cite{Appl1, Appl2}. This is a wide bandgap, unintentionally doped n-type semiconductor that is known for its high optical transparency and low electrical resistance \cite{Ref3}.
The possibility of fabricating SnO$_2$ aerogels further increases its applicability in space applications \cite{Ref1} and as a thermal insulator \cite{Ref2} because of high surface areas, low density, and tunable electrical conductivity \cite{SnO2Gen,GelTheory}. However, the performance of SnO$_2$ aerogels in these applications is often limited by surface defects, which can significantly alter their electronic structure and optical properties \cite{Def1}.

In this study, we investigate the impact of surface defects on the electronic properties of SnO$_2$ aerogels, with a particular focus on their influence on the bandgap. Our research reveals that the presence of surface defects leads to an apparent increase in the bandgap, potentially affecting the material's performance. Bulk SnO$_2$ exhibits an optical bandgap energy ($E_g$) of 3.6 eV, while SnO$_2$ in aerogel form shows an apparent $E_g$  around 4.6 eV, which significantly limits its electrical conductivity due to the reduced number of available charge carriers \cite{bandgap1, bandgap2, Ref4}. Surface defects, particularly dangling bonds on Sn-rich surfaces, lead to a Burstein-Moss shift \cite{Def2,Def3,Ref5}. This phenomenon elevates the Fermi level, resulting in both an apparent increase in the bandgap ($E_g$) and the formation of additional surface states, which significantly affect the material's electronic behavior \cite{shift1, shift2,Ref6}.
To address this issue, we explore different strategies for defect passivation, aiming to optimize the electronic properties of SnO$_2$ aerogels while maintaining its unique structural characteristics.

Initially, we examined a conventional annealing approach as a means to passivate surface defects \cite{anneal1,anneal2}. However, our findings demonstrate that this method, while traditionally effective in reducing defects, ultimately compromises its structure \cite{Ref4}. On the other hand, we demonstrate that aging the hydrogel phase in hydrogen peroxide (H$_2$O$_2$) results in a significant reduction of the apparent bandgap. Crucially, this strategy maintains the structural integrity of the material, overcoming the limitations observed with traditional annealing techniques.

Our work provides new insights into the nature of surface defects in high-surface area SnO$_2$ aerogels and offers a promising route for their mitigation. The findings presented here have the potential to enhance the performance of SnO$_2$ aerogels in various applications, paving the way for their improved use in next-generation devices and technologies. 

The structure of this paper is the following. Next, we discuss the materials and methods used to fabricate and measure the material. We then present the results obtained using Fourier transform infrared spectroscopy (FTIR), Williamson-Hall (WH) analysis performed on X-ray diffraction data (XRD), and ultraviolet-visible spectroscopy (UV-Vis).

\section{Materials and Methods}
\begin{figure}[H]
    \centering
    \includegraphics[ width=.5\textwidth]{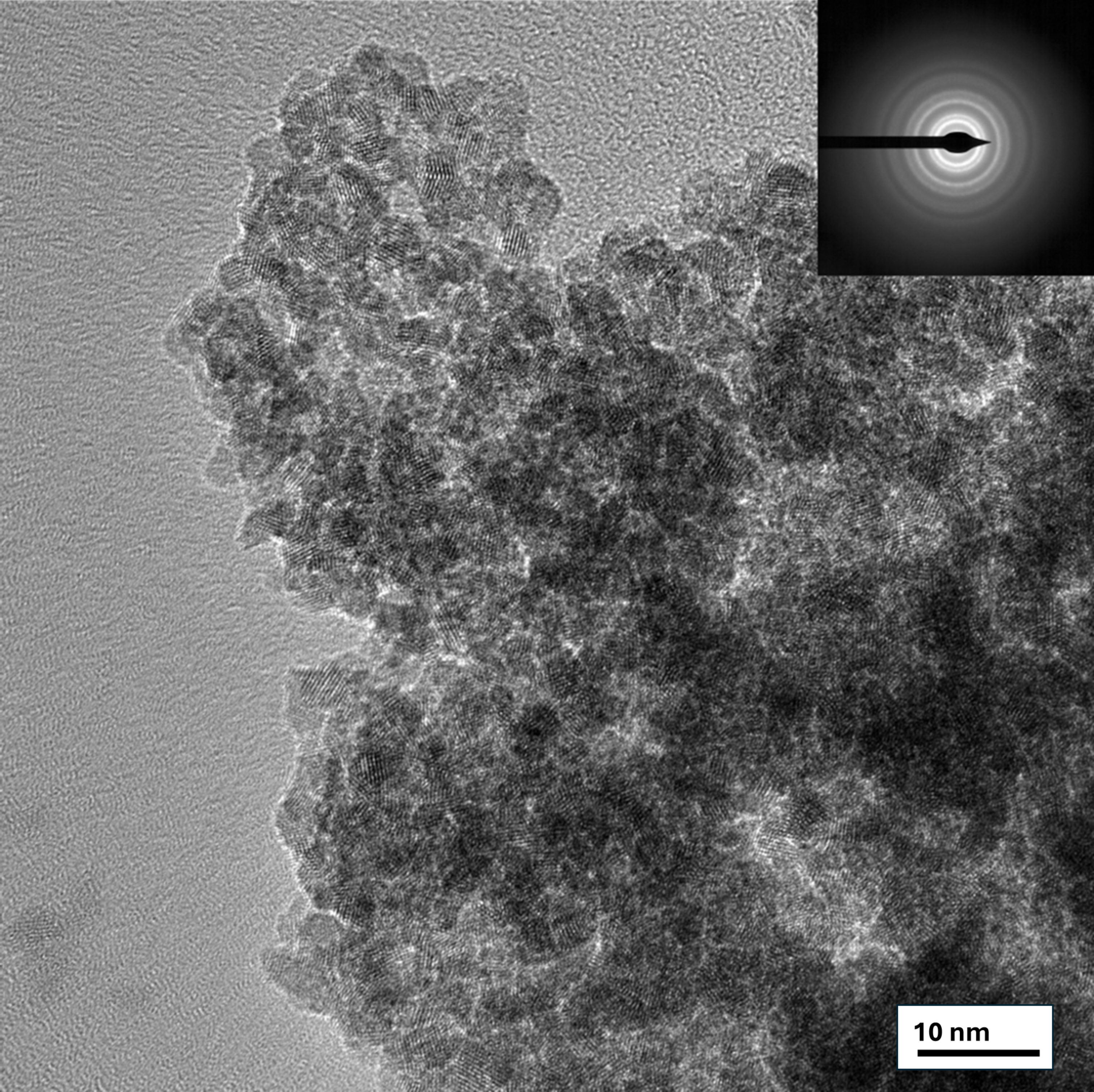}
    \caption{HRTEM image of the nanocrystallite array of the unannealed aerogel structure. A collection of nanocrystallites with an average diameter of $2.53$ nm is visible in the image.}
    \label{fig:1}
\end{figure}

All the chemicals used in the synthesis process were purchased from Sigma Aldrich and were used as received without further purification. An epoxide method was used to synthesize the SnO$_2$ aerogels \cite{Ref7}. In a typical synthesis, 0.65 g of tin tetrachloride pentahydrate (SnCl$_4$$\cdot$5H$_2$O) was dissolved in a 1:2 co-solvent mixture of ethanol (EtOH) and water (H$_2$O) under magnetic stirring for 2 minutes. After the complete dissolution of the tin precursor, 0.8 ml of propylene oxide (CH$_3$CHCH$_2$O) was gradually added to the solution. The mixture was immediately poured into a silicone mold, initiating hydrogel formation within 40 seconds. The hydrogel was aged in 200-proof ethanol for 24 hours, followed by a solvent exchange. Subsequently, a critical point drying process using liquid CO$_2$ was employed to convert the hydrogel into an aerogel using a Pelco CPD-2 critical point dryer. Figure \ref{fig:1} shows a high-resolution transmission electron microscope (HRTEM) image of the formed aerogel using a Talos F200i TEM. 
\noindent From the HRTEM image of the unannealed aerogel, the average crystallite size was calculated to be $2.53$ nm, and the rings in the selected area electron diffraction (SAED) pattern reveal that the SnO$_2$ aerogel is polycrystalline. By measuring the radius of the rings, the interplanar spacing can be calculated as \cite{CameraEq,BraggsLaw}:
\begin{equation}
    d_{hkl} = \frac{L\lambda}{R},
     \label{eq:z}
\end{equation}
where $d_{hkl}$ is the interplanar spacing of the crystal planes with Miller indices ($hkl$), $L$ is the distance from the specimen to the detector, $\lambda$ is the electron wavelength, and $R$ is the radius of the diffraction ring. SnO$_2$ has a tetragonal crystal rutile structure, and the $hkl$ indices are calculated as \cite{tetraRutile}:
\begin{equation}
    \frac{1}{d_{hkl}^2} = \frac{h^2+k^2}{a^2}+\frac{l^2}{c^2},
     \label{eq:y}
\end{equation}
where $a = 4.836$\text{\AA} and $c = 3.2742$\text{\AA} are the lattice parameters of the tetragonal structure \cite{LatticeParam}. From the diffraction pattern of the HRTEM image, the crystal planes most prominent are $(110)$, $(101)$, $(211)$, $(112)$, $(212)$, and $(322)$.

Once the aerogel was formed, the material underwent annealing at atmospheric pressure in a tube furnace at temperatures of $100^\circ$C, $200^\circ$C, $300^\circ$C, and $400^\circ$C for $30$ minutes each to study the effects of annealing on the crystalline structure of the SnO$_2$ aerogels. During the annealing procedure, the samples were exposed to the oxygen from the environment to passivate possible dangling bonds. Additionally, to passivate the dangling bonds on the material's surface without annealing, a tin oxide hydrogel sample was aged for $24$ hours in a $3$\% H$_2$O$_2$:H$_2$O solution.

FTIR (Bruker Alpha Spectrometer) was utilized to analyze the chemical composition of the aerogel and identify any potential contaminants or unreacted precursors. Each sample was fragmented into smaller pieces for analysis.   

A Rigaku Miniflex 6G with an ASC-8 powder sample attachment was used to perform XRD measurements on the powdered samples. The peaks in the diffractograms obtained from each sample were fitted with Gaussian curves to obtain peak broadenings. To calculate the crystallite size, the Debye-Scherrer equation was used \cite{Ref8}:
\begin{equation}
    D = \frac{K\lambda}{B_{hkl}cos(\theta)}.
     \label{eq:1}
\end{equation}
In this equation, $D$ is the crystallite size and $K$ is Sherrer's constant. We assumed $K=0.94$ for spherical crystallites  with cubic symmetry \cite{Ref9}.  $\lambda=1.54056$ \text{\AA} is the incident wavelength, $B_{hkl}$ is the full-width at half maximum (FWHM) of the peaks, and $\theta$ is the peak position for the corresponding $hkl$ plane. With Eq. \ref{eq:1}, the Williamson-Hall method was implemented to find the strain of the crystallites. This was performed using the Williamson-Hall equation \cite{Ref10}: 
\begin{equation}
    \beta\\cos(\theta) = C\epsilon\\sin(\theta)+\frac{K\lambda}{D}.
    \label{eq:2}
\end{equation}
We used $C=4$ as a constant related to the train and strain distribution \cite{Ref11,Ref12}. Also in the equation $\epsilon$ is the strain, $K$ is Sherrer's constant, $\lambda$ is the incident wavelength, and $D$ is the crystallite size. The strain was calculated with the slope of a linear regression of sin($\theta$) as a function of $\beta\cdot$cos($\theta$) for each sample.

UV-Vis was performed using an Agilent Cary 60 UV-Vis spectrophotometer. A reference of ethanol in a cuvette was taken, and then powdered aerogel was added to the ethanol for measurement to find the $E_g$ and the Urbach energy ($E_u$) \cite{Urbach}. The Urbach energy represents the width of the tail of localized states in the bandgap, extending from the band edges into the forbidden gap. A larger Urbach energy indicates a higher degree of disorder in the material. This disorder can be due to various factors such as impurities or structural defects.
The Urbach energy was obtained from the tail equation \cite{Ref16}:
\begin{equation}
    \alpha = \alpha_0 exp\left(\frac{h\nu-E_g}{E_u}\right),
    \label{eq:5}
\end{equation}
where $\alpha_0$ is the absorption constant, $h$ is Plank's constant, $\nu$ is the photon wavelength, $E_g$ is the optical bandgap energy, and $E_u$ is the Urbach energy. By taking the natural logarithm of the absorption coefficient and plotting this data against photon energy ($h\nu$), a linear fit was applied to the absorption data in the Urbach tail region. The $E_u$ was then derived from the inverse of the slope.

In optical experiments, Beer-Lambert's law was used to calculate the absorbance expressed by the equation \cite{Ref13}: 
\begin{equation}
    \alpha = log_{10}\left(\frac{I_o}{I}\right),
    \label{eq:3}
\end{equation}
where $I_o$ is the incident light, $I$ is the transmitted light to the spectrophotometer, and $\alpha$ is the absorbance. The absorbance $\alpha$ was later used in Tauc's equation to calculate $E_g$ \cite{Ref14}:
\begin{equation}
    (\alpha  h  \nu)^{1/\gamma} = \alpha_0(h \nu - E_g),
    \label{eq:4}
\end{equation}
where $h$ is the Plank's constant, $\nu$ is the photon's frequency, and $\gamma$ is the factor determined by the nature of the electronic transition involved. In the study of SnO$_2$ aerogels, $1/\gamma$ is assumed to be $2$ corresponding to a direct transition, and $\alpha_0$ is the energy-independent constant \cite{Ref15}. 

\section{Results}
\begin{figure}[H]
    \centering
    \includegraphics[ width=.6\textwidth]{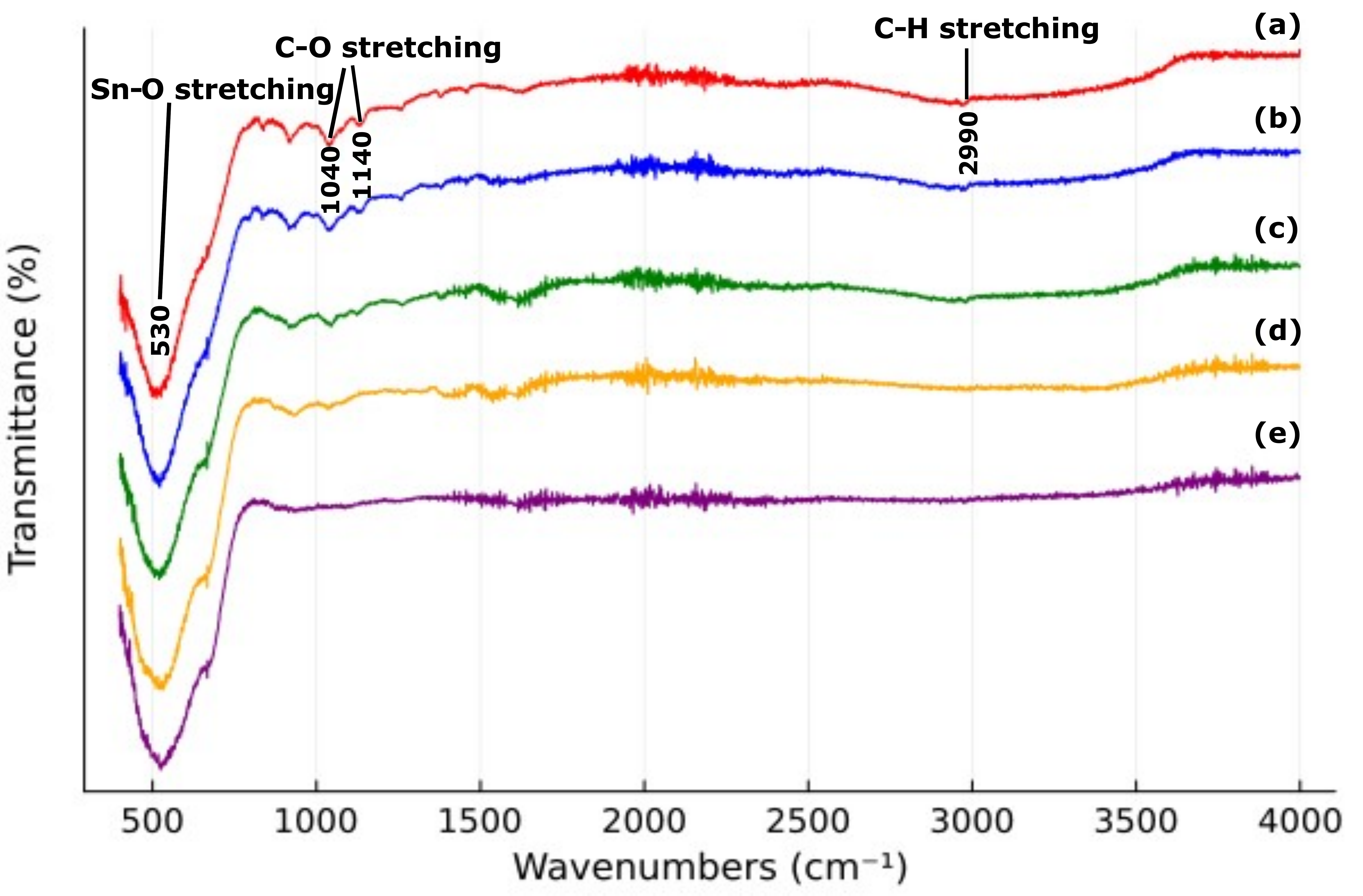}
    \caption{FTIR spectrum of the unannealed sample (a) and the annealed samples at 100$^\circ$C (b), 200$^\circ$C (c), 300$^\circ$C (d), and 400$^\circ$C (e).}
    \label{fig:2}
\end{figure}

The FTIR spectrum of the samples in Fig. \ref{fig:2} reveals distinct peaks corresponding to C-H stretching at $2990$--$3000$ cm$^{-1}$, C-O stretching at $1040$--$1050$ cm$^{-1}$ and $1130$--$1140$ cm$^{-1}$. The large feature at $530$ cm$^{-1}$ can be attributed to Sn-O stretching \cite{Ref4}. 

\begin{figure}[H]
    \centering
    \includegraphics[ width=1\textwidth]{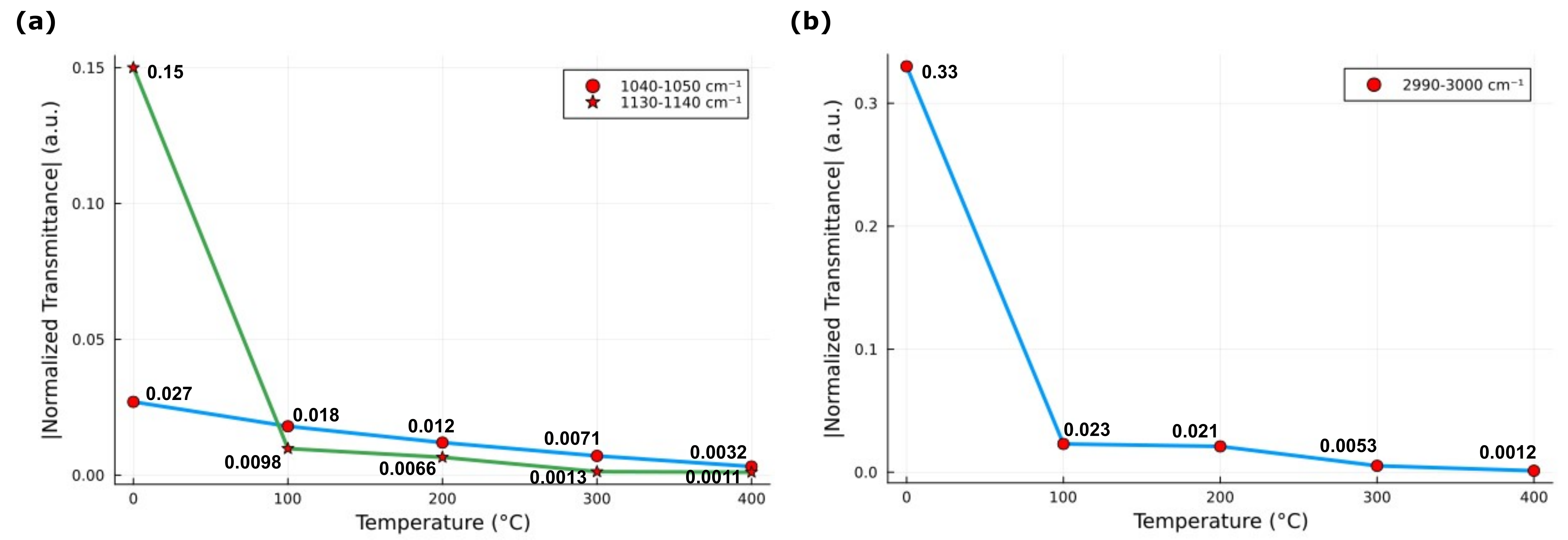}
    \caption{Transmission intensities corresponding to: a) C-O stretching, and b) C-H stretching.}
    \label{fig:3}
\end{figure}

\noindent The obtained C-O and C-H stretching bands can be attributed to residual impurities containing carbon and hydroxyl groups. The influence of annealing on  impurities is illustrated in Figure \ref{fig:3},
which depicts the transmittance corresponding to impurity peaks at varying temperatures.  
 \noindent The graphs were normalized to a baseline of zero to visualize the relative changes in transmittance. Each data point represents the magnitude of transmittance of each peak height, and the reduction in transmittance suggests that fewer impurities are present at higher temperature. Figure \ref{fig:3} demonstrates a trend in both C-O and C-H stretching which indicates that impurities are reduced as the annealing process progresses.

\begin{figure}[H]
    \centering
    \includegraphics[ width=1\textwidth]{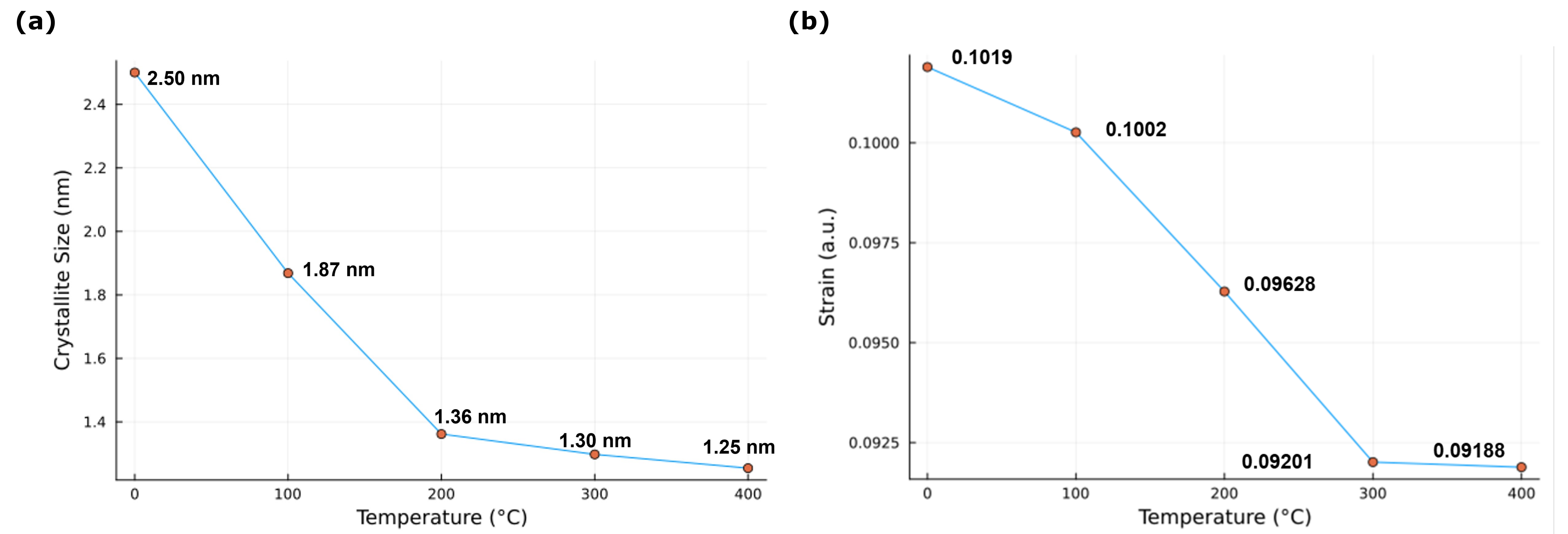}
    \caption{a) Crystallite sizes calculated with Debye-Scherrer equation, and b) the strain of the crystal calculated with the Williamson-Hall equation.}
    \label{fig:4}
\end{figure}

The XRD results for the Debye-Scherrer equation, Eq. \ref{eq:1}, for each temperature is illustrated in Fig. \ref{fig:4}(a). Through this method, the average crystallite sizes are calculated at each temperature. The average crystallite size of the unannealed sample, measured using XRD, was found to be $2.5$ nm, which is in close agreement with the value of $2.53$ nm obtained from electron diffraction.

\noindent By applying the Debye-Scherrer equation, the Williamson-Hall method (Eq. \ref{eq:2}) was implemented to analyze the strain of the crystallites. These results are shown in Fig. \ref{fig:4}(b). The crystallite size exhibits a near linear decrease from room temperature to $200^{\circ}$, with a rate of shrinkage of approximately $6$ pm$/^{\circ}$C. Beyond $200^{\circ}$, the rate of shrinkage significantly diminishes, reaching approximately $0.55$ pm$/^{\circ}$C. Strain reduces approximately $5.5$\% from room temperature up to $300^{\circ}$.

\begin{figure}[H]
    \centering
    \includegraphics[ width=1\textwidth]{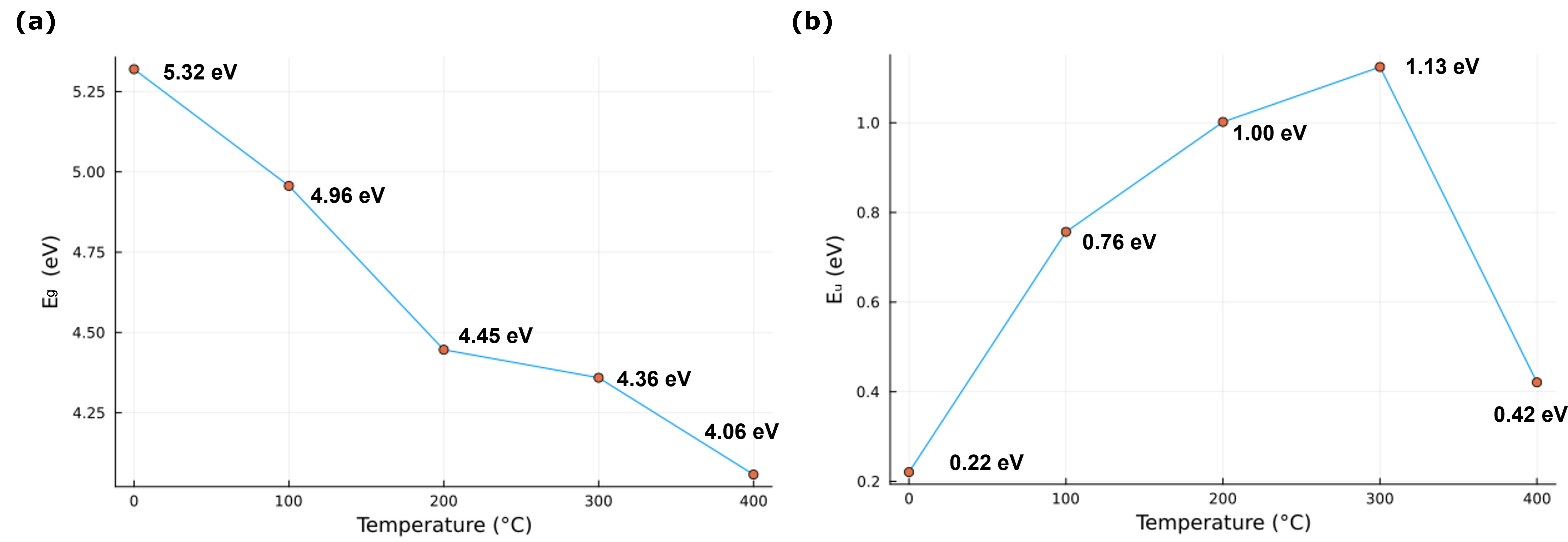}
    \caption{a) Bandgap energies ($E_g$) as a function of the annealing temperature, and b) the corresponding Urbach energy ($E_u$).}
    \label{fig:5}
\end{figure}

The apparent bandgaps obtained from applying Eq. \ref{eq:4} on Tauc plots are shown in Fig. \ref{fig:5}(a), while the corresponding Urbach energies are shown in Fig. \ref{fig:5}(b). The bandgap decreases at a nearly linear rate of approximately $-3.15$ meV/$^{\circ}$ from room temperature to $400^{\circ}$. The Urbach energy has a four-fold increase between room temperature to $300^{\circ}$ but then drops by $63$ \% at $400^{\circ}$. This drop at $300^{\circ}$ is consistent with our measurements of strain and aligns with the expected phase transition from SnO$_2$ to SnO that begins at approximately $300^{\circ}$ \cite{PhaseT}.
 
The optical properties of the unannealed, peroxide-passivated samples are detailed in Table \ref{table:1}.

\begin{table}[H]
    \centering
    \begin{tabular}{|c|c|c|}
        \hline
          & \textbf{Control} & \textbf{H$_2$O$_2$} \\ \hline
        \textbf{$E_u$} & 0.2207 eV & 0.147 eV \\ \hline
        \textbf{$E_g$} & 5.32 eV   & 3.64 eV \\ \hline
    \end{tabular}
    \caption{Urbach energy ($E_u$) and bandgap energy ($E_g$) for control unannealed and peroxide-passivated samples.}
    \label{table:1}
\end{table}
\noindent Since there were no annealing treatments performed on the peroxied-passivated sample, the XRD and FTIR results are comparable to the control unannealed sample.

\section{Discussion}
\begin{figure}[H]
    \centering
    \includegraphics[ width=.7\textwidth]{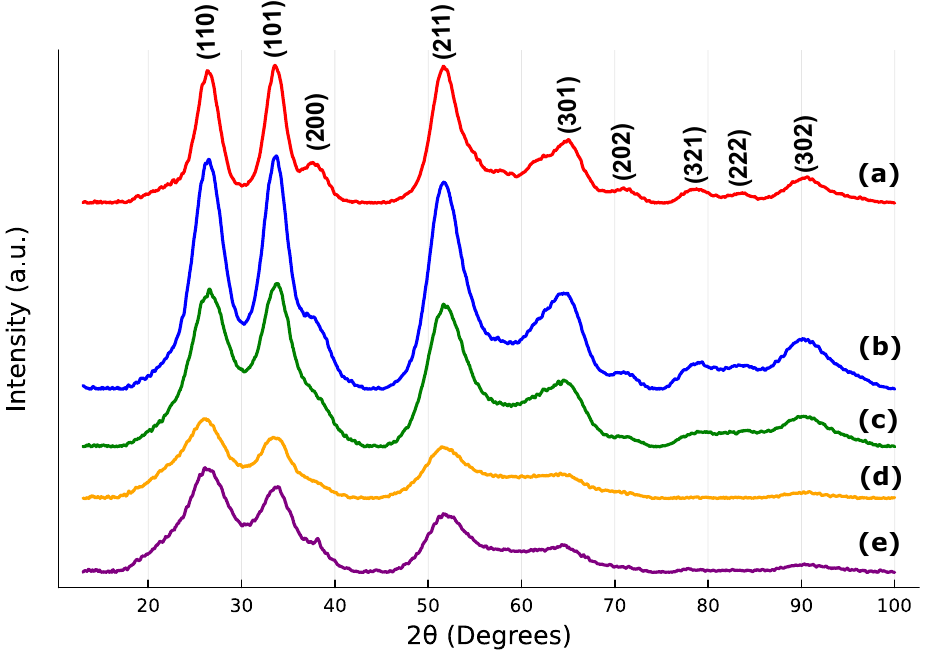}
    \caption{XRD patterns of the a) unannealed sample and the annealed samples at b) $100^\circ$C, c) $200^\circ$C, d) $300^\circ$C (d), and e) $400^\circ$C.}
    \label{fig:6}
\end{figure}

While annealing removes the impurities of the sample at each temperature as shown explicitly in Fig. \ref{fig:3}, the positive effect of annealing is due to a partial phase change that occurs at around 300$^\circ$C \cite{PhaseT}. In the XRD peaks shown in Fig. \ref{fig:6}, some peaks begin to broaden and disappear at 300$^\circ$C. Around 400$^\circ$C, a new peak begins to emerge at 2$\theta$ = 37$^\circ$ which corresponds to the $(002)$ planes of SnO \cite{Ref17}.

This indicates there is a partial phase change from SnO$_2$ to SnO, and this reduction is due to oxygen loss during annealing \cite{Ref18}. This behavior aligns with findings from previous literature where it was demonstrated that thermal activation of metal-oxide aerogels leads to the generation of oxygen vacancies \cite{Ref19}. 
We have previously shown through photoluminescence measurements that oxygen vacancies and oxygen complexes such as (Sn$_i^{+4}+$V$_0$)$^0$ are prevalent in sol-gel fabricated SnO2 hydrogels \cite{Ref4}. These defects not only facilitate the phase change but also significantly influence the material's electronic properties by increasing the number of electron trapping sites.  
 This partial phase change would explain why the crystallite size decreases in the results of the Debye-Scherrer equation shown in Fig. \ref{fig:4}(a). It can be concluded that the oxygen loss causes the SnO$_2$ crystallites to become thermodynamically unstable, and they break down into smaller SnO crystallites that are more stable at the given annealing temperature. This partial phase change can also be seen in the $E_u$ depicted in Fig. \ref{fig:5}(b) as well. In the unannealed sample, the disorder is quantified by an Urbach energy of $0.22$ eV which suggests that there are fewer localized states within the bandgap, and as more oxygen vacancies are created more localized states and disorder is seen within the material. At $300^\circ$C, the highest state of disorder is seen with the $E_u$ reaching $1.13$ eV. Once the material stabilizes at $400^\circ$C the $E_u$ decreases to $0.42$ eV, indicating a more ordered state, likely due to the phase change stabilizing the material.  The formation of SnO is favored at higher temperatures, where the energy required for oxygen vacancy formation is compensated by the release of heat during the reduction. The cause for this partial phase change can be attributed to the negative enthalpy of the formation of SnO \cite{Enthalpies}, indicating that the system achieves greater stability as it loses oxygen. Even with the creation of the oxygen vacancies and the partial phase change, Fig. \ref{fig:4}(b) shows that as the SnO phase forms, the material enters a more stable, relaxed state and annealing further reduces strain by relieving structural defects and grain boundary stresses. The reduction in the $E_g$ can also be attributed to the partial phase change since SnO has a narrower $E_g$ than SnO$_2$. Moreover, after the phase becomes more stable at $400^\circ$C, there is still an apparent $E_g$ of $4.06$ eV while SnO has a  $E_g$ typically within the range of $1.75$--$3.4$ eV \cite{Ref20,Ref21}. When the unannealed sample is compared to the sample treated with H$_2$O$_2$, the $E_u$ is reduced from $0.2207$ eV to $0.147$ eV (a $33$ \% improvement), and the $E_g$ is also significantly reduced from $5.32$ eV to $3.64$ eV which is close to the expected $E_g$ of SnO$_2$. This strongly supports the idea thesis the apparent $E_g$ is heavily influenced by the dangling bonds on the surface of the material. H$_2$O$_2$ is a strong oxidizing agent, and when it interacts with reactive dangling bonds, the H$_2$O$_2$ donates oxygen to the surface, filling the dangling bonds and passivating the surface. Similar effects were demonstrated in studies of H$_2$O$_2$ on Ge($100$), where the dangling bonds were significantly reduced compared to H$_2$O passivation \cite{Ref22}.

\section{Conclusions}
In this study, SnO$_2$ aerogels were synthesized using an epoxide technique, and the effects of post-synthesis annealing and H$_2$O$_2$ passivation on surface defects were investigated to optimize the materials electronic properties for potential use in electronic applications. SnO$_2$ aerogel was shown to exhibit an increased apparent bandgap due to surface defects when compared to the typical bandgap of bulk SnO$_2$. Annealing the aerogels induces a partial phase change from SnO$_2$ to SnO, which creates oxygen vacancies, reduces strain in the material, and leads to a moderate improvement in the materials properties. This phase change does not fully address surface defects, but passivation with H$_2$O$_2$ effectively reduces both the bandgap and disorder in the material without compromising structural integrity, suggesting that surface dangling bonds significantly contribute to the increased bandgap. In conclusion, while annealing has some benefits, H$_2$O$_2$ passivation offers a more efficient way to enhance the electronic properties of SnO$_2$ aerogels by minimizing surface defects and lowering the E$_g$. These findings provide new insights into the potential of SnO$_2$ aerogels for electronic applications such as solar cells and sensors.

\section{Acknowledgements}
The research was sponsored by the Army Research Laboratory and was accomplished under Cooperative Agreement Number W911NF-23-2-0014. The views and conclusions contained in this document are those of the authors and should not be interpreted as representing the official policies, either expressed or implied, of the Army Research Laboratory or the U.S. Government. The U.S. Government is authorized to reproduce and distribute reprints for Government purposes notwithstanding any copyright notation herein.

This material is also based upon work supported by the National Science Foundation under Grant No. 2425226.

\section{Conflict of Interest}
The authors declare that they have no conflict of interest.

\newpage
\singlespacing

\bibliography{main} 
\bibliographystyle{unsrt}

\end{document}